\documentclass[10pt,journal]{IEEEtran}
\pdfoutput=1
\usepackage{cite}
\ifCLASSINFOpdf
   \usepackage[pdftex]{graphicx}
   \graphicspath{{../pdf/}{../jpeg/}}
   \DeclareGraphicsExtensions{.pdf,.jpeg,.png}
\else
\fi
\usepackage{amsmath}
\usepackage{algorithm}
\usepackage{algpseudocode}
\usepackage{array}
\usepackage{mdwmath}
\usepackage{mdwtab}
\usepackage{eqparbox}
\ifCLASSOPTIONcompsoc
  \usepackage[caption=false,font=normalsize,labelfont=sf,textfont=sf]{subfig}
\else
  \usepackage[caption=false,font=footnotesize]{subfig}
\fi
 \usepackage{dblfloatfix}
\usepackage{url}
\newcommand\MYhyperrefoptions{bookmarks=true,bookmarksnumbered=true,
pdfpagemode={UseOutlines},plainpages=false,pdfpagelabels=true,
colorlinks=true,linkcolor={black},citecolor={black},urlcolor={black},
pdftitle={Content Delivery Latency of Caching Strategies for Information-Centric IoT},%
pdfsubject={Information-Centric Network},%
pdfauthor={Jakob Pfender, Alvin Valera, Winston K.G. Seah},%
pdfkeywords={Information-Centric Networking, Named Data Networking, Internet of Things, In-Network Caching, Network Topology}}%
\ifCLASSINFOpdf
\usepackage[\MYhyperrefoptions,pdftex]{hyperref}
\else
\usepackage[\MYhyperrefoptions,breaklinks=true,dvips]{hyperref}
\usepackage{breakurl}
\fi
\usepackage{booktabs} % For formal tables

\usepackage[acronym]{glossaries}
\usepackage{hyperref}
\usepackage[capitalise]{cleveref}
\crefformat{figure}{Fig.~#2#1#3}
\usepackage{tabularx}
\usepackage{tabu}
\usepackage{subfig}
\usepackage{xcolor}

\usepackage{tikz}
\usepackage{forest}

\usepackage{flushend}

\usetikzlibrary{positioning}

\definecolor{mynew}{HTML}{136F63}
\definecolor{myneww}{HTML}{1C7C54}
\definecolor{mynewww}{HTML}{EF2D56}
\definecolor{myrev}{HTML}{9097C0}

\newcommand{\ws}[1]{\textcolor{blue}{#1}}
\renewcommand{\ws}[1]{#1}
\newcommand{\wsq}[1]{\textcolor{red}{#1}}
\renewcommand{\wsq}[1]{#1}

\newcommand{\new}[1]{\textcolor{mynew}{#1}}
\renewcommand{\new}[1]{#1}
\newcommand{\neww}[1]{\textcolor{myneww}{#1}}
\renewcommand{\neww}[1]{#1}
\newcommand{\newww}[1]{\textcolor{mynewww}{#1}}
\renewcommand{\newww}[1]{#1}
\newcommand{\rev}[1]{\textcolor{myrev}{#1}}
\renewcommand{\rev}[1]{#1}

\setacronymstyle{long-short}

\newacronym{AS}{AS}{Autonomous System}
\newacronym{CCN}{CCN}{Content-Centric Networking}
\newacronym{CEE}{CEE}{Cache Everything Everywhere}
\newacronym{CRR}{CRR}{Cache Retention Ratio}
\newacronym{CS}{CS}{Content Store}
\newacronym{DM}{DM}{Diversity Metric}
\newacronym{FIB}{FIB}{Forwarding Information Base}
\newacronym{ICN}{ICN}{In\-for\-ma\-tion-Cen\-tric Networking}
\newacronym{IoT}{IoT}{Internet of Things}
\newacronym{LCD}{LCD}{Leave Copy Down}
\newacronym{LCE}{LCE}{Leave Copy Everywhere}
\newacronym{LFU}{LFU}{Least Frequently Used}
\newacronym{LRU}{LRU}{Least Recently Used}
\newacronym{MCD}{MCD}{Move Copy Down}
\newacronym{MDMR}{MDMR}{Max Diversity Most Recent}
\newacronym{NDN}{NDN}{Named Data Networking}
\newacronym{RR}{RR}{Random Replacement}
\newacronym{SNC}{SNC}{Selective Neighbour Caching}
\newacronym{TSB}{TSB}{Time Since Birth}
\newacronym{TSI}{TSI}{Time Since Inception}
\newacronym{WSN}{WSN}{wireless sensor network}

\forestset{%
my forest/.style={%
    for tree={%
            node options={text width=3.5cm, align=center},
            l sep=1cm, 
            parent anchor=south, 
            child anchor=north,
            if n children=0{tier=word}{},
        edge path={%
            \noexpand\path [\forestoption{edge}] (!u.parent anchor) -- +(0,-15pt) -| (.child anchor)\forestoption{edge label};
        },
    }
}
}

\definecolor{prodc}{HTML}{0A2239}
\definecolor{conc}{HTML}{53A2BE}
\definecolor{nodec}{HTML}{1D84B5}
\definecolor{cachec}{HTML}{132E32}

\tikzstyle{producers}=[draw, circle, fill=prodc, minimum width=20pt]
\tikzstyle{consumers}=[draw, circle, fill=conc, minimum width=10pt]
\tikzstyle{routers}=[draw, circle, fill=nodec, minimum width=10pt]
\tikzstyle{caches}=[draw, circle, fill=cachec, minimum width=20pt]

\begin{document}
%\title{On the Latency Effects of Caching in Information-Centric IoT}
\title{Content Delivery Latency of Caching Strategies\\ for Information-Centric IoT}

\author{Jakob~Pfender,~\IEEEmembership{Student~Member,~IEEE,}
        Alvin~Valera,~\IEEEmembership{Member,~IEEE,}
        and~Winston~K.~G.~Seah,~\IEEEmembership{Senior Member,~IEEE}% <-this % stops a space
\thanks{J. Pfender, A. Valera,
  and W. Seah are with the School of Engineering and Computer Science,
  Victoria University of Wellington, New Zealand.
  E-mail: \{jakob.pfender, alvin.valera, winston.seah\}@ecs.vuw.ac.nz}% <-this % stops a space
\thanks{Manuscript received Month xx, xxxx; revised Month xx, xxxx.}}


\markboth{ARXIV}%
{Pfender \MakeLowercase{\textit{et al.}}: Content Delivery Latency of Caching Strategies for Information-Centric IoT}

\IEEEpubid{0000--0000/00\$00.00~\copyright~2019 IEEE}


\IEEEtitleabstractindextext{%
\begin{abstract}
  In-network caching is a central aspect of \gls*{ICN}. It enables the
  rapid distribution of content across the network, alleviating strain
  on content producers and reducing content delivery latencies.
  \gls*{ICN} has emerged as a promising candidate for use in the
  \gls*{IoT}. However, \gls*{IoT} devices operate under severe
  constraints, most notably limited memory. This means that nodes cannot
  indiscriminately cache all content; instead, there is a need for
  a caching strategy that decides what content to cache. Furthermore,
  many applications in the \gls*{IoT} space are time-sensitive;
  therefore, finding a caching strategy that minimises the latency
  between content request and delivery is desirable. In this paper, we
  evaluate a number of \gls*{ICN} caching strategies in regards to
  latency and hop count reduction using \gls*{IoT} devices in a physical
  testbed. We find that the topology of the network, and thus the
  routing algorithm used to generate forwarding information, has
  a significant impact on the performance of a given caching strategy.
  To the best of our knowledge, this is the first study that focuses on
  latency effects in ICN-IoT caching while using real IoT hardware, and
  the first to explicitly discuss the link between routing algorithm,
  network topology, and caching effects.
\end{abstract}

}

\maketitle

% is not selected <OR> if conference mode is selected - because compsoc
% conference papers position the abstract like regular (non-compsoc)
\IEEEdisplaynontitleabstractindextext

% \fi
\IEEEpeerreviewmaketitle

%!TEX root = paper.tex
%==============================================================================

\ifCLASSOPTIONcompsoc{}
\IEEEraisesectionheading{\section{Introduction}\label{sec:introduction}}
\else
\section{Introduction}%
\label{sec:introduction}
\fi

\glsreset{ICN}
\glsreset{IoT}
\IEEEPARstart{T}{he} \new{content-centric nature and slim network stack
of \gls*{ICN} make it an ideal candidate for a future network
architecture for \gls*{IoT}
applications~\cite{adhatarao_isi:_2018, liu_object-oriented_2017,
yan_novel_2014, arshad_hierarchical_2018, arshad_towards_2017}. However,
the fact that resources such as memory, processing power, and battery
life are traditionally severely limited in the \gls*{IoT} means that the
automatic and indiscriminate caching of all content in the network, as
is common in traditional \gls*{ICN}, can not be transferred to this
domain. Memory in particular is a much more valuable resource in
\gls*{IoT} applications, so approaches that simply assume that all
content will be cached at all nodes are not feasible. Instead, the
questions of what content should be cached, which nodes should cache it,
and for how long it should be cached become central questions of ICN-IoT
in-network caching. Although these questions have been explored in
traditional \gls*{ICN} research, a large number of potential solutions
still assume cache sizes that are unrealistic in \gls*{IoT}
environments, and their application to \gls*{IoT}-specific deployments
has proven sub-optimal~\cite{hail_caching_2015, hail_performance_2015,
amadeo_information-centric_2016, arshad_information-centric_2017,
arshad_recent_2018}. It is therefore desirable to find caching solutions
that are especially geared towards addressing the constraints unique to
the \gls*{IoT}.}

Apart from increasing content redundancy and decreasing network load,
\new{efficient} in-network caching can also reduce content delivery
latencies by making content more readily available across the
network~\cite{LCN2015:LAC}. Operating under the assumption that content,
once produced, will remain useful for a certain amount of time and will
be requested by multiple consumers during its lifetime, an effective
caching strategy will minimise the distance between consumers and cached
copies of the content they require. Given an information-centric
\gls*{IoT} application where a reduction in content delivery latency is
the primary objective, we want to compare and contrast the effect
different approaches to \gls*{ICN} in-network caching have on this
metric.

\new{The question of where in the network content should be cached is
one of the most defining problems of \neww{in-network} caching
research~\cite{carofiglio_experimental_2011, psaras_probabilistic_2012,
chai_cache_2013, zhu_caching_2018}. In particular, it has not been
definitively established whether it is better to cache content towards
the edge of the network (i.e. closer to the consumer) or towards the
core, i.e.\ closer to the producer.} Intuitively, caching closer to the
consumer \new{would seem} to make more sense; after all, if we keep
content close to the producer, we can reduce \new{the} load \new{on that
particular node}, but the potential latency improvements seem to be
minimal. If content is requested from a particular region of the
network, it is probably safe to assume that it will be requested from
that region again in the future. If the content was already cached close
to that region, retrieving it will be much faster. This is the argument
for caching towards the consumer and there are a number of caching
strategies that employ this paradigm~\cite{psaras_probabilistic_2012,
fayazbakhsh_less_2013, wang_faircache:_2016, meddeb_cache_2018}.
However, as we will show in \cref{sub:topology}, depending on topology,
there is also a strong argument for the inverse approach.

\IEEEpubidadjcol
In this study, we will be evaluating \neww{a number of different
in-network caching} strategies. To achieve this, we design an experiment
using real \gls*{IoT} hardware in a physical testbed meant to
\ws{emulate} the conditions of a typical \gls*{IoT} application as
closely as possible. We measure content delivery latencies as well as
the average reduction in hop count between cached content and original
storage location. \new{We also investigate the extent to which the
effectiveness of a given strategy is influenced by the network topology
and thus the routing algorithm.}

\new{Although other authors have investigated and compared ICN-IoT
caching strategies using various performance metrics (see
\cref{sec:related_work}), there has not been an in-depth study
exclusively focused on latency effects, and none of the existing studies
have used physical IoT hardware to perform their experiments.
Furthermore, no existing studies explicitly take the strong correlation
between routing algorithm, network topology, and caching effects into
account}. These are the key contributions of this paper.

This \ws{paper} is structured as follows:

\begin{itemize}
    \item We introduce a number of modern caching strategies for
        \newww{\gls*{ICN}} and discuss their suitability for the
        \gls*{IoT} (\cref{sec:caching_strategies}).
    \item We evaluate and compare the caching strategies introduced in
        \cref{sec:caching_strategies} in regards to metrics such as hop
        count and latency reduction (\cref{sec:evaluation}). \new{We
        also discuss the effects of network topology.}
    \item We discuss related research in the field of caching in
        \gls*{ICN} with a focus on \gls*{IoT} (\cref{sec:related_work})
    \item Finally, we present potential future research directions
        (\cref{sec:conclusion}).
\end{itemize}

\section{Caching Strategies}%
\label{sec:caching_strategies}

\glsreset{CCN}
The original \gls*{CCN} proposal~\cite{jacobson_networking_2009} did not
place a strong focus on caching policies. It was assumed that nodes
would simply cache all received content indiscriminately. It has since
been shown~\cite{psaras_modelling_2011, carofiglio_experimental_2011,
psaras_probabilistic_2012, zhang_cache_2019} that caches can be utilised
more effectively using more advanced caching policies.

In this section, we will introduce different approaches to caching
policy with varying complexity. \new{For every strategy that we evaluate
in \cref{sec:evaluation}, we will provide a pseudocode definition here.
Every strategy provides two functions, \textsc{handle\_interest()} and
\textsc{handle\_data()}, which define what the strategy does upon
reception of an Interest or Data packet, respectively. There are also
functions that are not further defined in the pseudocode, such as
\texttt{canSatisfy()}, which returns \texttt{true} if the incoming
Interest can be satisfied locally and \texttt{false} otherwise;
\texttt{getData()}, which retrieves a content chunk from the local
\gls*{CS}; and the NDN primitives \texttt{reply()} (for replying to an
Interest with a Data packet), \texttt{forward()} (for forwarding
Interest and Data packets to the next hop), and \texttt{cache()} (for
caching content).}

\subsection{Cache Everything Everywhere (CEE)}%
\label{sub:cee}

\begin{algorithm}[!t]
  \caption{Cache Everything Everywhere (CEE)}\label{alg:cee}
  \begin{algorithmic}[1]
    \Function{handle\_interest}{Interest}
        \If{canSatisfy(Interest)}
            \State Data $\gets$ getData(Interest)
            \State reply(Data)
        \Else
            \State forward(Interest)
        \EndIf
    \EndFunction
    \State
    \Function{handle\_data}{Data}
        \State cache(Data)
        \State forward(Data)
    \EndFunction
  \end{algorithmic}
\end{algorithm}

The simplest approach to the caching decision is for every node to cache
every piece of incoming content. This requires no computational overhead
and leads to rapid proliferation of content across the network; however,
it also results in high cache redundancy and thus suboptimal resource
use. It has become consensus~\cite{psaras_modelling_2011,
carofiglio_experimental_2011, psaras_probabilistic_2012} that this
strategy is not optimal in terms of providing redundant access to as
large a subsection of available content as possible; instead, caches
tend to hold only the most recent content and the danger of thrashing is
high. Depending on the application, however, \emph{\gls*{CEE}}
(\cref{alg:cee}) can have a positive effect on content delivery
latencies\ws{,} since it is the fastest way to proliferate new content
throughout the network\ws{;} the more content requests are skewed
towards recent content, the more they may benefit from \gls*{CEE}.

\subsection{Leave Copy Down (LCD)}%
\label{sub:lcd}

\begin{algorithm}[!b]
  \caption{Leave Copy Down (LCD)}%
  \label{alg:lcd}
  \begin{algorithmic}[1]
    \Function{handle\_interest}{Interest}
        \If{canSatisfy(Interest)}
            \State Data $\gets$ getData(Interest)
            \State Data.TSB $\gets 1$
            \State reply(Data)
        \Else
            \State forward(Interest)
        \EndIf
    \EndFunction
    \State
    \Function{handle\_data}{Data}
        \If{Data.TSB $= 1$}
            \State cache(Data)
        \EndIf
        \State Data.TSB $\gets$ Data.TSB $+ 1$
        \State forward(Data)
    \EndFunction
  \end{algorithmic}
\end{algorithm}

If we do not want to cache everything at every node, but also do not
want to introduce complex additional steps into the caching process, we
can use a policy called \emph{\gls*{LCD}} (\cref{alg:lcd}). In
\gls*{LCD}, content is always cached only at the next hop from the node
where the cache hit occurred (i.e.\ the first node to receive the Data)
and nowhere else. \new{We achieve this by extending the Data packet with
a \gls*{TSB} field, which counts the number of hops since the creation
(``birth'') of the Data packet. \gls*{TSB} is initially 1 and is
incremented every time the Data packet is forwarded. Nodes only cache
Data with a \gls*{TSB} of exactly 1.}

\new{\gls*{LCD}} is a somewhat ``conservative'' caching strategy that
tends to keep content close to the producer, but still alleviates load
on the producer\rev{. Since every cache hit results in the content being
cached %in the next hop,
one hop closer to the requesting consumer,} popular content that is requested with high
frequency will gradually move the content closer to consumers with each
step. 

\subsection{Probabilistic Caching}%
\label{sub:probabilistic_caching}

\begin{algorithm}[!t]
    \caption{Prob($p$)}%
    \label{alg:prob05}
    \begin{algorithmic}[1]
        \Function{handle\_interest}{Interest} 
            \If{canSatisfy(Interest)}
                \State Data $\gets$ getData(Interest)
                \State reply(Data)
            \Else
                \State forward(Interest)
            \EndIf
        \EndFunction
        \State
        \Function{handle\_data}{Data}
            \If{rand() $< p$}
                \State cache(Data)
            \EndIf
            \State forward(Data)
        \EndFunction
    \end{algorithmic}
\end{algorithm}

The easiest way to achieve higher cache diversity without increasing the
complexity of the caching policy is to cache probabilistically. In the
static version of this approach, commonly known as $Prob(p)$
(\cref{alg:prob05}), we set a static probability $p$ that governs how
likely a given node will cache a given content chunk. It has been
shown~\cite{zhang_survey_2015, hail_caching_2015} that $Prob(p)$
outperforms \gls*{CEE} in terms of cache diversity, and that lower
values for $p$ correlate with higher
diversity~\cite{psaras_probabilistic_2012, tarnoi_performance_2014,
hail_performance_2015, hail_caching_2015, zhang_survey_2015,
arshad_information-centric_2017}.

However, instead of defining an \emph{a priori} caching probability that
is the same for every caching decision at every node, we can also design
a technique that dynamically computes a caching probability for each
individual node or even for each content chunk, based on available
information, in order to adapt the caching behaviour to the state of the
network. These strategies could be based purely on node-local
information, such as the current contents of the cache or the node's
battery levels; they could \ws{also} be based on properties of the incoming
content chunk, such as its age, type, or producer; or they could be
based on information from the wider network, such as the position of the
caching node in the network topology or the cache contents of
neighbouring nodes.

\begin{algorithm}[!t]
    \caption{ProbCache}%
    \label{alg:probcache}
    \begin{algorithmic}[1]
        \Function{handle\_interest}{Interest} 
            \If{canSatisfy(Interest)}
                \State Data $\gets$ getData(Interest)
                \State Data.TSI $\gets$ Interest.TSI
                \State Data.TSB $\gets 1$
                \State reply(Data)
            \Else
                \State Interest.TSI $\gets$ Interest.TSI $+ 1$
                \State forward(Interest)
            \EndIf
        \EndFunction
        \State
        \Function{handle\_data}{Data}
            \State Data.TSB $\gets$ Data.TSB $+ 1$
            \State CacheWeight = Data.TSB / Data.TSI
            \If{rand() $<$ CacheWeight}
                \State cache(Data)
            \EndIf
            \State forward(Data)
        \EndFunction
    \end{algorithmic}
\end{algorithm}

Psaras~\emph{et~al}.\ propose
\emph{ProbCache}~\cite{psaras_probabilistic_2012}
(\cref{alg:probcache}), which computes the caching probability of
a given content chunk based on the total number of hops between its
producer and the consumer that requested it, as well as the number of
hops remaining \new{on the path} to the consumer. For a given content
chunk travelling a path between producer and consumer, ProbCache
determines the \new{\emph{cache weight} at each node, which is
determined by the Data packet's \gls*{TSB} (see \cref{sub:lcd}) as well
as its \gls*{TSI}, which is the number of hops between the creation
(``inception'') of the corresponding Interest packet and the cache hit,
i.e.\ the total length of the path between consumer and producer. This
means} that content chunks are cached with a higher probability towards
the edges of the network (i.e., closer to the consumer), adjusted by the
length of the \new{Interest packet's path}. The authors find that in
traditional \gls*{ICN}, ProbCache increases the cache hit ratio, reduces
the average number of hops required to hit requested content, and
reduces the number of cache evictions.

As \neww{will be discussed in more detail in \cref{sub:topology}},
depending on network topology, caching closer to the consumer may not
necessarily be the most efficient caching policy\rev{; in some
topologies, caching closer to the producer is more beneficial}.
ProbCache is easily modified to take this consideration into account by
simply inverting the caching probability such that content is more
likely to be cached near the producer. We call this modified strategy
\emph{ProbCache-Inv} and it is identical to ProbCache in every way
except that the final caching probability is inverted (i.e.\ \newww{line
16} in \cref{alg:probcache} reads \texttt{if rand() < (1 - CacheWeight)}).
 
\subsection{Cooperative Caching}%
\label{sub:cooperative_caching}

Cooperative caching is an umbrella term for caching strategies that take
more than local information into account, i.e., strategies in which
nodes either \emph{implicitly} or \emph{explicitly} coordinate with
their neighbours to ensure optimal caching. In explicit coordination,
nodes may exchange information about their cache contents and/or the
contents they have received on a periodic or ad hoc basis in order to
make caching decisions, or even forward content chunks to one another
for caching.

\begin{algorithm}[!t]
  \caption{Labels($k$)}%
    \label{alg:labels}
    \begin{algorithmic}[1]
        \Function{handle\_interest}{Interest} 
            \If{canSatisfy(Interest)}
                \State Data $\gets$ getData(Interest)
                \State reply(Data)
            \Else
                \State forward(Interest)
            \EndIf
        \EndFunction
        \State
        \Function{handle\_data}{Data}
            \If{Data.ID mod $k =$ myLabel}
                \State cache(Data)
            \EndIf
            \State forward(Data)
        \EndFunction
    \end{algorithmic}
\end{algorithm}

In implicit coordination, nodes follow \emph{a priori} rules that govern
what content they can cache\rev{, thus avoiding the need for explicit
coordination}. One example of this was proposed by Li and
Simon~\cite{li_time-shifted_2011} \new{(\cref{alg:labels}),} \ws{where}
each node is assigned a fixed label $l < k$ \rev{at setup} and only
caches content chunks whose IDs modulo $k$ are equal to $l$. This
ensures \rev{that} cached content \rev{is automatically stratified into
equal subsets and evenly distributed across the network} without the
overhead of explicit coordination between nodes. By adjusting $k$, it is
possible to control the level of stratification. We will call this
caching strategy \emph{Labels} for the remainder of this study.

\begin{algorithm}[!b]
  \caption{Intervals($i$)}%
  \label{alg:intervals}
  \begin{algorithmic}[1]
    \Function{handle\_interest}{Interest}
        \If{canSatisfy(Interest)}
            \State Data $\gets$ getData(Interest)
            \State Data.Interval $\gets i$
            \State reply(Data)
        \Else
            \State forward(Interest)
        \EndIf
    \EndFunction
    \State
    \Function{handle\_data}{Data}
        \If{Data.Interval $= 0$}
            \State cache(Data)
            \State Data.Interval $\gets i$
        \Else
            \State Data.Interval $\gets$ Data.Interval $- 1$
        \EndIf
        \State forward(Data)
    \EndFunction
  \end{algorithmic}
\end{algorithm}

Zeng and Hong~\cite{zeng_caching_2011} \new{(\cref{alg:intervals})}
propose an implicitly coordinated caching strategy that uses hop
distance to determine the caching decision. Data packets are extended by
a \new{pre-determined} \emph{data interval} value \new{$i$}. Each node
along the path decrements this value by 1 when forwarding the packet. If
a node decrements its value to 0, the packet is cached at that node and
the data interval is reset \new{to $i$}. This ensures that data are
implicitly cached at regular distances from producers without requiring
any topological information or coordination. We will call this caching
strategy \emph{Intervals} for the remainder of this study.

\subsection{Other approaches}
\label{sub:caching_decision_other}

\new{In the following, we will briefly cover a number of alternative
strategies which we will not be evaluating in this study because their
complexity, their overhead, or other factors make them infeasible for
use in the \gls*{IoT}.}

\new{\gls*{MCD} is a variant of \gls*{LCD} (\cref{sub:lcd}) in which
instead of simply copying a content chunk to the next node whenever
a cache hit occurs, that content chunk is explicitly \emph{moved} to the
next node; i.e., it is deleted from the current cache and only stored in
the next cache. This would free up cache space for new content near the
core.} \rev{We decided not to investigate this approach because the
underlying principle is virtually identical to \gls*{LCD} and there is
not much further insight to be gained from studying it.}

\new{ProbCache (\cref{sub:probabilistic_caching}) is representative of
a broader family of caching strategies with dynamic probability, which
can take a multitude of weighted factors into account when calculating
their caching probabilities. Other such approaches may use some
combination of node battery level and cache occupancy in their
calculations, along with some information about the incoming content,
\ws{e.g.} its freshness~\cite{hail_caching_2015,
doan_van_efficient_2018} or its popularity~\cite{chen_brr-cvr:_2016},
whether the content is already cached in a neighbouring
node~\cite{zhang_lf:_2015}, and/or topological information such as hop
count~\cite{doan_van_efficient_2018, zhang_lf:_2015} or the caching
node's centrality~\cite{naz_dynamic_2018, chen_brr-cvr:_2016}.} \rev{The
space of possible permutations of parameters for probabilistic caching
is very large, so we decided to focus on ProbCache as the sole
representative.}

\new{Explicit cache coordination strategies tend to be more complex than
their implicit counterparts (\cref{sub:cooperative_caching}), requiring
more communication among the nodes and more calculations to maintain
a consistent state. For example,
Liu~\textit{et~al}.~\cite{liu_novel_2016} propose a strategy that
coordinates nodes by constructing a virtual backbone network using
graph-theoretical concepts to create a node hierarchy with core nodes
responsible for caching. However, since one of the primary goals of
optimising strategies for use in the \gls*{IoT} should be a minimisation
of overhead, we will not be considering any explicit coordination
strategies in this comparison.}

\new{Vural~\textit{et~al}.~\cite{vural_caching_2017} propose a strategy that
uses the popularity of content classes to calculate the cost function
for the caching of incoming content. Nodes cache content until it
reaches a certain age, and for each incoming content chunk, the caching
node calculates how many Interests the chunk can be expected to satisfy
during its lifetime. Content that is expected to serve more Interests
--- because its content is more popular and/or because it is more fresh
--- is cached with a higher probability. Popularity is also used in
TCCN~\cite{song_tccn:_2015}, where content is enhanced with tags and
nodes keep track of the popularity of each tag as well as how often it
is cached in nearby nodes.} \rev{These popularity-based approaches are
quite demanding of node resources, especially memory, since all nodes
need to keep track of the global popularity of all content. In addition,
approaches that rely on neighbours sharing information to estimate
probability would incur an additional communication overhead. These
factors make this class of strategies infeasible for the \gls*{IoT}.}

\new{Chai~\emph{et~al}.~\cite{chai_cache_2013} propose
\emph{Betw}/\emph{EgoBetw}, a caching scheme that takes into account the
topology of the network, specifically the \emph{betweenness centrality}
of the caching node. Betweenness centrality measures how many times
a given node lies on the paths between all pairs of nodes in a given
topology. In Chai~\emph{et~al.}'s scheme, content is cached at the node
with the highest centrality on the delivery path. That way, content is
automatically stored at the most central locations in the network, where
Interests are most likely to result in cache hits. However, this
strategy requires a costly setup phase. As an alternative,
Chai~\emph{et~al.} propose the modified version \emph{EgoBetw}, which
requires no setup and instead relies on nodes exchanging connectivity
information with their neighbours to approximate their centrality. The
trade-off is increased communications and computational overhead during
operation, which is antithetical to what \gls*{IoT} algorithms try to
accomplish. Thus, it is highly doubtful whether \emph{Betw/EgoBetw} are
feasible for information-centric \gls*{IoT}, and we decided not to
implement these strategies for this study.}

\section{Evaluation}%
\label{sec:evaluation}

In this section, we present the main contribution of the paper:
a comprehensive comparison and evaluation of several caching strategies
for information-centric \gls*{ICN} with a focus on latency and hop
reduction. For this comparison, we ran a series of experiments on the
\textbf{FIT IoT-LAB}~\cite{adjih_fit_2015} open testbed. We used
IoT-LAB's specially developed \textbf{M3
node}\footnote{\url{https://github.com/iot-lab/iot-lab/wiki/Hardware_M3-node}}\hspace*{-1ex},
which has an STM32 (ARM Cortex M3) microcontroller and an Atmel
AT86RF231 2.4 GHz transceiver, as our \gls*{IoT} hardware. As firmware
for the nodes, we use a simple
\textbf{RIOT-OS}~\cite{baccelli_riot:_2018} application using
\textbf{CCN-lite}\footnote{https://github.com/cn-uofbasel/ccn-lite} as
the \gls*{ICN} implementation, modified to support the different caching
strategies. 

The experiments were conducted on the \emph{Grenoble}
site\footnote{\url{https://www.iot-lab.info/deployment/grenoble/}} of
the IoT-LAB testbed. The site features about 380 M3 nodes, which are
distributed across the rooms and corridors of one floor of an office
building. This means that nodes are subject to realistic conditions
found in indoor \gls*{IoT} deployments, such as multipath effects,
reflection, and absorption caused by walls, doors, and windows made of
various materials, as well as unpredictable interference by other
wireless signals and people moving around the building. These conditions
mean that data gathered will be very close to what might be expected in
a real-world deployment.

Of the 380 available nodes, each experiment run is conducted on an
arbitrary subset of 50 nodes, chosen randomly each time. \rev{This
ensures that the topology is different in each experiment run and also
that the nodes will not be too strongly connected due to having a large
number of one-hop neighbours; this is desirable as it allows us to study
the effects of unreliable connections more closely.} The transmission
range of individual nodes is not enough to reach all other nodes in the
building, so communication will be predominantly multihop. \rev{In
a typical topology generated by this random selection of nodes, the}
mean path length is between 3 and 4 hops and the maximum is 6 hops.
\new{Cache sizes are kept intentionally small; each node's cache can
store up to 5 unique content chunks (all content chunks have the same
size).}

The experiment is managed by a control script using the IoT-LAB API,
which has access to all node caches, outputs, and inputs. An experiment
begins with a brief (30 seconds) setup phase, in which every node
advertises its own prefix (dictated by its address), which is then
propagated through the rest of the network using
\textbf{HoPP}'s~\cite{gundogan_hopp:_2018} routing algorithm. \new{HoPP
is primarily a publish-and-subscribe scheme for information-centric IoT,
but also includes a prefix advertisement mechanism based on the
\textbf{Trickle}~\cite{RFC6206} algorithm. The fact that the routing
algorithm is based on Trickle} also means that nodes' \glspl*{FIB}
can be kept up to date during runtime. After setup is complete, every
node will request a piece of content with a random ID in $\left\{0,
\ldots, 49\right\}$ from each of the prefixes in its \gls*{FIB}.
Interest and Data packets are handled as specified by the \gls*{NDN}
standard. The first time a node receives an Interest for a content chunk
it owns, it produces that content chunk (the actual payload is
irrelevant for our experiment) and sends it back towards the consumer.
Caching of content chunks at intermediate nodes is dictated by the
caching strategy selected for the experiment.

\new{The network topology is a direct result of the \gls*{FIB} contents,
which in turn are a direct result of the routing algorithm. In the
\neww{HoPP/Trickle} routing algorithm, prefix advertisements are
propagated in a tree-like fashion. A producer will advertise its own
prefixes with a rank of 0, which is then increased by each node that
forwards the advertisement. When forwarding interests, nodes will always
prefer the \gls*{FIB} entry with the lowest rank.}

The controller takes periodic snapshots of cache contents and
\glspl*{FIB} and logs statistics such as latency and hop counts. We use
this information to evaluate the caching strategies in the rest of this
section.

\begin{figure*}[!t]
  \begin{minipage}[b]{0.32\textwidth}
    \centering
    \includegraphics[trim=0mm 5mm 0mm 5mm, width=\columnwidth]{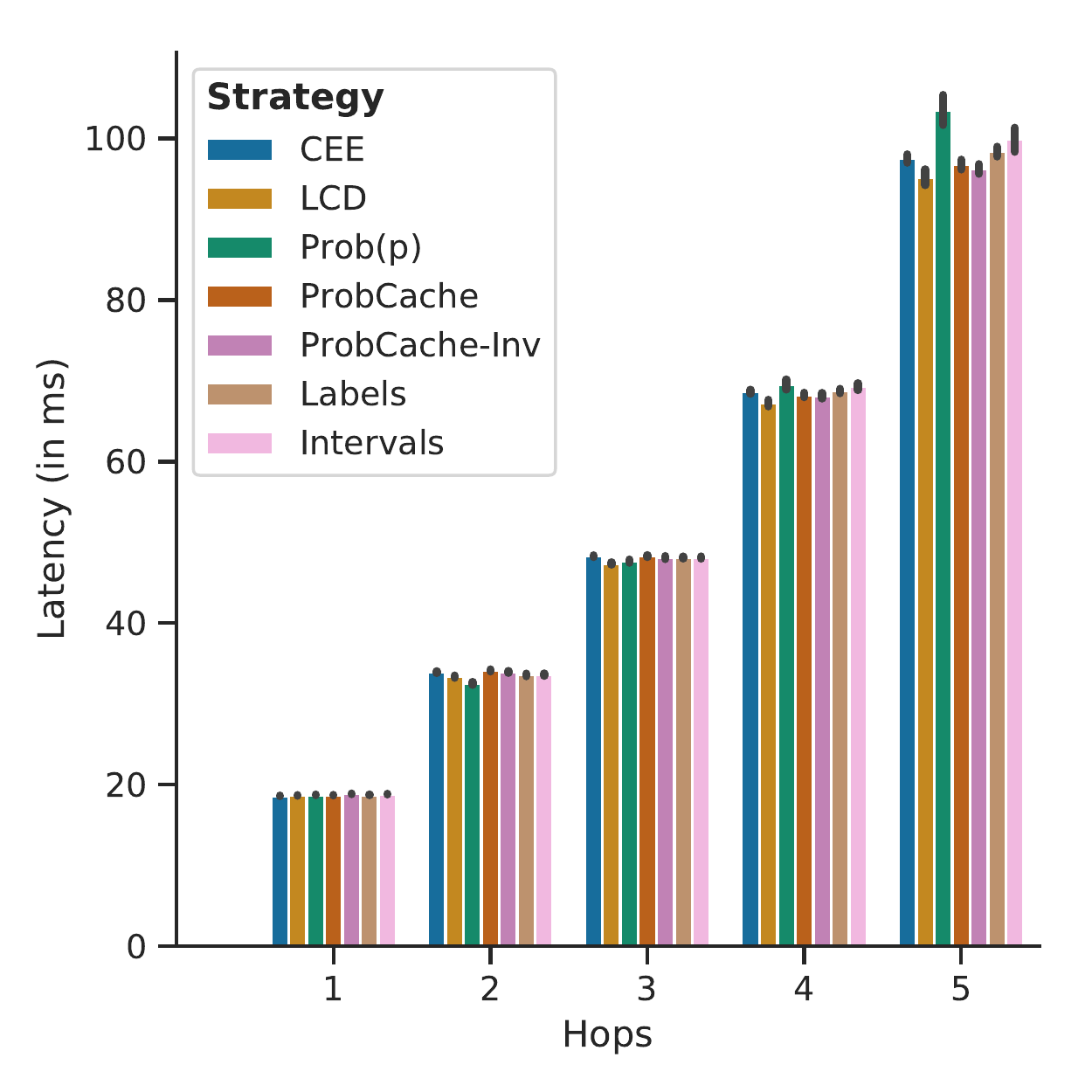}
    \caption{Relation between hop count and latency}%
    \label{fig:hops_latency}
  \end{minipage}
  \hfil
  \begin{minipage}[b]{0.32\textwidth}
    \centering
    \begin{tikzpicture}[thick, scale=0.9, every node/.style={transform shape}]
        \node[producers] at (0,0) (prod) {}; 

        \node[routers, above of=prod] (a) {};
        \node[routers, above right of=prod] (b) {};
        \node[routers, right of=prod] (c) {};
        \node[routers, below right of=prod] (d) {};
        \node[routers, below of=prod] (e) {};
        \node[routers, below left of=prod] (f) {};
        \node[routers, left of=prod] (g) {};
        \node[routers, above left of=prod] (h) {};

        \node[routers, above of=a] (i) {};
        \node[routers, above right of=b] (j) {};
        \node[routers, right of=c] (k) {};
        \node[routers, below right of=d] (l) {};
        \node[routers, below of=e] (m) {};
        \node[routers, below left of=f] (n) {};
        \node[routers, left of=g] (o) {};
        \node[routers, above left of=h] (p) {};

        \path (prod) edge (a);
        \path (prod) edge (b);
        \path (prod) edge (c);
        \path (prod) edge (d);
        \path (prod) edge (e);
        \path (prod) edge (f);
        \path (prod) edge (g);
        \path (prod) edge (h);

        \begin{scope} [dotted]
          \path (a) edge (i);
          \path (b) edge (j);
          \path (c) edge (k);
          \path (d) edge (l);
          \path (e) edge (m);
          \path (f) edge (n);
          \path (g) edge (o);
          \path (h) edge (p);
        \end{scope}

        \foreach \pos/\i in {above left of/1, above right of/2, above of/3}
        \node[consumers, \pos =i] (i\i) {};
        \foreach \speer/\peer in {i/i1,i/i2,i/i3}
        \path (\speer) edge (\peer);
        \foreach \pos/\i in {above of/1, right of/2, above right of/3}
        \node[consumers, \pos =j] (j\i) {};
        \foreach \speer/\peer in {j/j1,j/j2,j/j3}
        \path (\speer) edge (\peer);
        \foreach \pos/\i in {above right of/1, below right of/2, right of/3}
        \node[consumers, \pos =k] (k\i) {};
        \foreach \speer/\peer in {k/k1,k/k2,k/k3}
        \path (\speer) edge (\peer);
        \foreach \pos/\i in {right of/1, below of/2, below right of/3}
        \node[consumers, \pos =l] (l\i) {};
        \foreach \speer/\peer in {l/l1,l/l2,l/l3}
        \path (\speer) edge (\peer);
        \foreach \pos/\i in {below right of/1, below left of/2, below of/3}
        \node[consumers, \pos =m] (m\i) {};
        \foreach \speer/\peer in {m/m1,m/m2,m/m3}
        \path (\speer) edge (\peer);
        \foreach \pos/\i in {below of/1, left of/2, below left of/3}
        \node[consumers, \pos =n] (n\i) {};
        \foreach \speer/\peer in {n/n1,n/n2,n/n3}
        \path (\speer) edge (\peer);
        \foreach \pos/\i in {below left of/1, above left of/2, left of/3}
        \node[consumers, \pos =o] (o\i) {};
        \foreach \speer/\peer in {o/o1,o/o2,o/o3}
        \path (\speer) edge (\peer);
        \foreach \pos/\i in {left of/1, above of/2, above left of/3}
        \node[consumers, \pos =p] (p\i) {};
        \foreach \speer/\peer in {p/p1,p/p2,p/p3}
        \path (\speer) edge (\peer);
    \end{tikzpicture}
    \caption{Edge topology}
    \label{fig:edge_topology}
  \end{minipage}
  \hfil
  \begin{minipage}[b]{0.32\textwidth}
    \centering
    \begin{tikzpicture}[thick, scale=0.9, every node/.style={transform shape}]
        \node[producers] at (0,0) (prod) {}; 

        \node[routers, above of=prod] (a) {};
        \node[routers, right of=prod] (b) {};
        \node[routers, below of=prod] (c) {};
        \node[routers, left  of=prod] (d) {};

        \path (prod) edge (a);
        \path (prod) edge (b);
        \path (prod) edge (c);
        \path (prod) edge (d);

        \foreach \pos/\i in {above left of/1, above of/2, above right of/3}
        \node[routers, \pos =a] (a\i) {};
        \foreach \speer/\peer in {a/a1,a/a2,a/a3}
        \path (\speer) edge (\peer);
        \foreach \pos/\i in {above right of/1, right of/2, below right of/3}
        \node[routers, \pos =b] (b\i) {};
        \foreach \speer/\peer in {b/b1,b/b2,b/b3}
        \path (\speer) edge (\peer);
        \foreach \pos/\i in {below right of/1, below of/2, below left of/3}
        \node[routers, \pos =c] (c\i) {};
        \foreach \speer/\peer in {c/c1,c/c2,c/c3}
        \path (\speer) edge (\peer);
        \foreach \pos/\i in {below left of/1, left of/2, above left of/3}
        \node[routers, \pos =d] (d\i) {};
        \foreach \speer/\peer in {d/d1,d/d2,d/d3}
        \path (\speer) edge (\peer);

        \node[consumers, above left of=a1] (a1c) {};
        \node[consumers, above of=a2] (a2c) {};
        \node[consumers, above right of=a3] (a3c) {};
        \node[consumers, above right of=b1] (b1c) {};
        \node[consumers, right of=b2] (b2c) {};
        \node[consumers, below right of=b3] (b3c) {};
        \node[consumers, below right of=c1] (c1c) {};
        \node[consumers, below of=c2] (c2c) {};
        \node[consumers, below left of=c3] (c3c) {};
        \node[consumers, below left of=d1] (d1c) {};
        \node[consumers, left of=d2] (d2c) {};
        \node[consumers, above left of=d3] (d3c) {};

        \begin{scope} [dotted]
          \path (a1) edge (a1c);
          \path (a2) edge (a2c);
          \path (a3) edge (a3c);
          \path (b1) edge (b1c);
          \path (b2) edge (b2c);
          \path (b3) edge (b3c);
          \path (c1) edge (c1c);
          \path (c2) edge (c2c);
          \path (c3) edge (c3c);
          \path (d1) edge (d1c);
          \path (d2) edge (d2c);
          \path (d3) edge (d3c);
        \end{scope}
    \end{tikzpicture}
    \caption{Core topology}
    \label{fig:core_topology}
  \end{minipage}
\end{figure*}

\begin{figure*}[!t]
\begin{minipage}[b]{0.32\textwidth}
    \centering
    \includegraphics[width=\columnwidth]{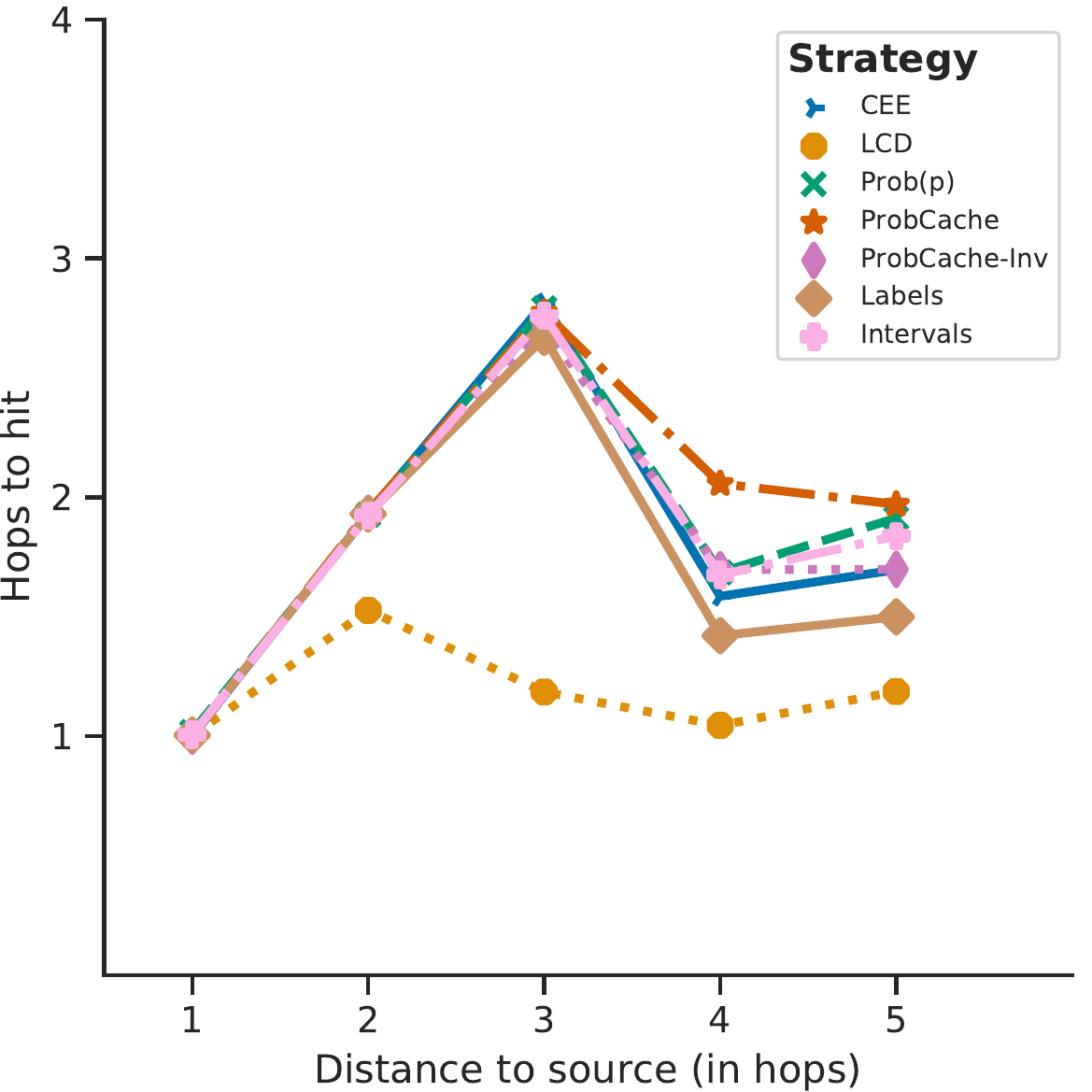}
\caption{Hop count reduction}%
\label{fig:distance_hops}
\end{minipage}
\hfil
\begin{minipage}[b]{0.32\textwidth}
    \centering
    \includegraphics[width=\columnwidth]{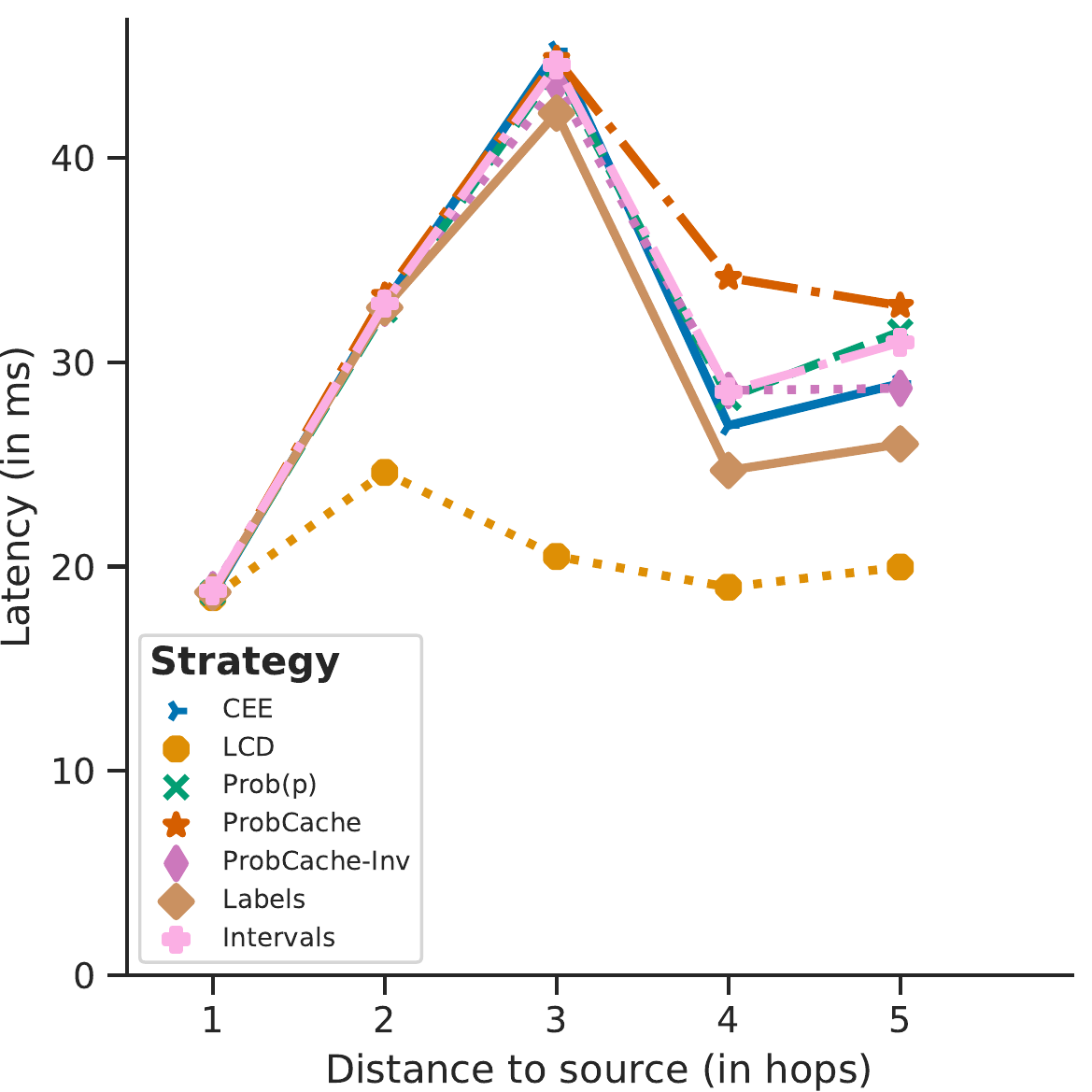}
\caption{Latency by distance}%
\label{fig:distance_latency}
\end{minipage}
\hfil
\begin{minipage}[b]{0.32\textwidth}
    \centering
    \includegraphics[width=\columnwidth]{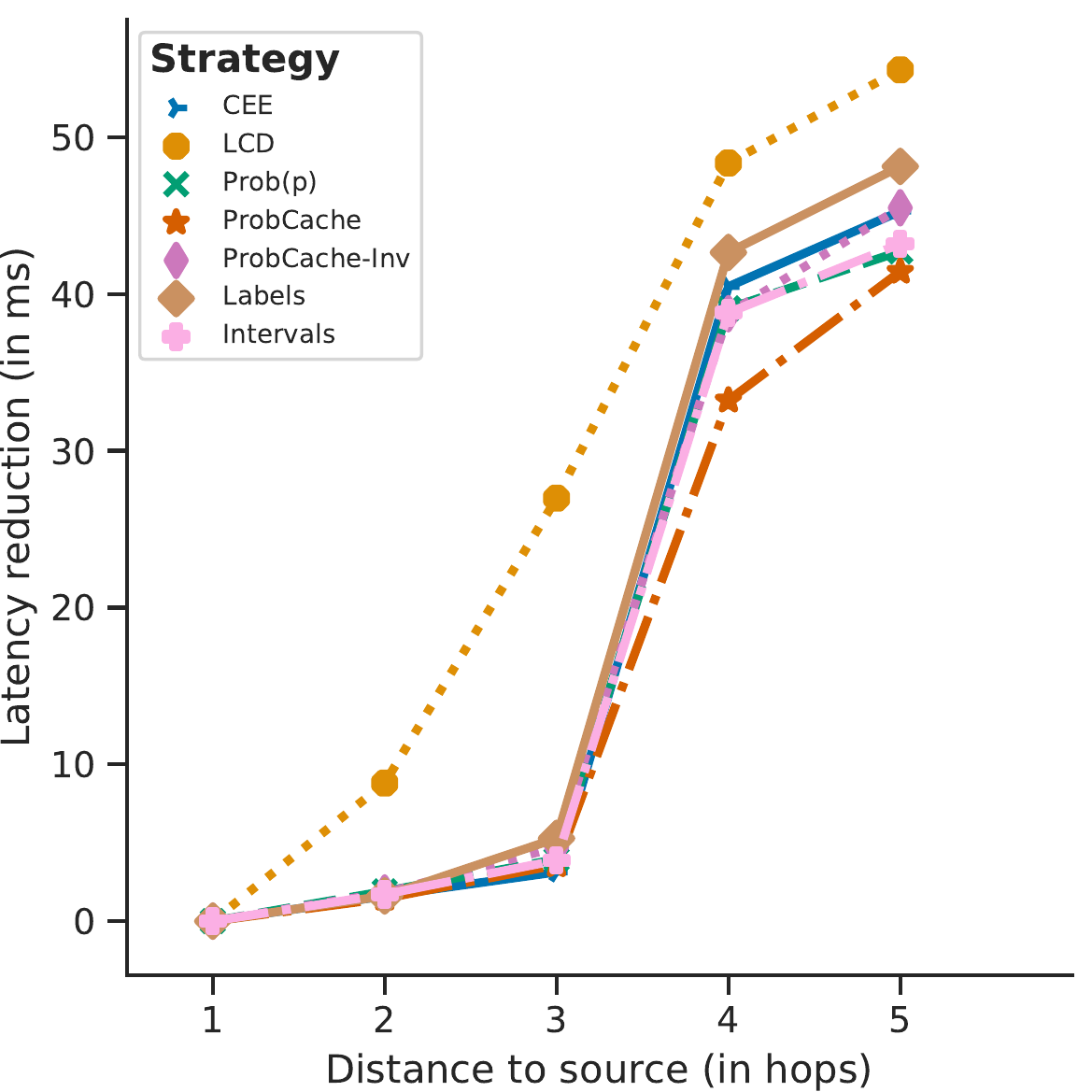}
\caption{Latency reduction compared to no caching}%
\label{fig:latency_reduction_cee}
\end{minipage}
\end{figure*}

\subsection{Relation between hop count and latency}%
\label{sub:hops_latency}
\Cref{fig:hops_latency} shows how the latency (\ws{specifically,} the
data retrieval delay\ws{, which is} the time between Interest generation
and satisfaction) is affected by the number of hops taken to retrieve
the content. This does not differentiate between cached content and
content produced by the prefix owner\ws{, i.e.} the hop count shown here
is the number of hops \newww{traversed by the Data to the requester from
either its original producer or from a caching node.} \rev{This means
that this figure shows only the correlation between hop count and
latency for individual transfers, which is linear because nodes are
evenly spaced in the network.} The caching strategy \rev{does} not have
an impact on this metric in terms of packet travel time, \rev{because
the caching strategy affects the hop count and not the actual
per-hop transfer speed.}

The only impact the caching strategy \new{could} have on this metric is
the computational overhead as each hop on the Data path needs to make
a decision. However, the difference appears negligible. \new{Thus}, we
can already conclude that the complexity of \new{all} caching strategies
is within acceptable bounds, making them worth considering. For the rest
of this paper, this graphic mostly serves to illustrate the ground truth
of what latency to expect depending on the hop count. It will be useful
for contrasting with the metrics we will discuss in \new{the rest of
this section, as it effectively also represents the latencies we would
expect if we did not perform any in-network caching at all}.

\subsection{Hop count reduction}%
\label{sub:distance_hops}
For each Interest, we \ws{denote the number of} hops between its origin
and the owner of the prefix it is requesting \ws{as the}
\new{\emph{distance to source}. In other words, this is the number of
hops the Interest/Data packet would always have to travel if there were
no caches in the network}. We then relate this \ws{distance} to the
actual number of hops taken by the Data packet on the way back. \new{We
call this \emph{hops to hit} as it denotes the number of hops it
actually took for the Interest to reach a cache hit.} The more efficient
a caching strategy, the more \new{content} will be available in a cache
closer to the consumer, leading to a lower \new{average hops to hit
value. The difference between the distance to source and the hops to hit
is what we denote as the \emph{hop count reduction}}.

\Cref{fig:distance_hops} shows the average hop count reduction for the
different caching strategies at different distances. The first obvious
effect we can see is that \new{most} of the strategies only show
a significant hop count reduction starting from a minimum distance to
source. \new{In all strategies except for \gls*{LCD}, there is
a slight reduction at 3 hops and then a substantial one at 4 hops.
The reason for this is that} at shorter distances, there is
less cache space between the producer and the consumer, which means
fewer opportunities for content to be cached on the path. This makes it
much more likely that a request will have to be routed all the way to
the prefix owner to be satisfied. \rev{After a distance of 4 hops, the
hops to hit will increase again as the distance to source increases.}
The ``turning point\ws{''} at which caching begins to have a noticeable
impact seems to lie \new{between 3 to} 4 hops for most strategies. After
this point, there is enough cache space on the path that content is
likely to be found at a closer node. \new{\gls*{LCD}, on the other hand,
already shows a noticeable hop reduction at a distance of 2 hops, which
gets even stronger as the distance increases.}

\new{Another strategy that stands out is Labels, which in terms of
hop count reduction is second only to \gls*{LCD}. This strategy takes
a very different approach from \gls*{LCD} in that it favours an even
distribution of content by stratifying it according to content and node
IDs~\cite{li_time-shifted_2011}. It appears that this approach ---
explicitly ensuring that each individual piece of content is available
in the same number of nodes across the network --- results in
a beneficial cache distribution.}

\newww{It has been observed multiple
times~\cite{carofiglio_experimental_2011, psaras_modelling_2011,
psaras_probabilistic_2012, pfender_performance_2018} that \gls*{CEE} is
not an optimal caching strategy for \gls*{ICN}, and this is supported by
our results. The reason is that \gls*{CEE} is vulnerable to
\emph{thrashing} effects when nodes are caching high volumes of diverse
data; the limited size of caches in \gls*{IoT} only exacerbates this
effect.}

The fact that \gls*{LCD} exhibits a significantly greater reduction in
hop count points to a phenomenon alluded to in \cref{sec:introduction}:
The topology of the network has a significant impact on the
effectiveness of the chosen caching strategy. This is explained in more
detail in \cref{sub:topology}:

\subsection{Topology effects}%
\label{sub:topology}

\neww{In \cref{sec:introduction}, we claimed that although caching
content closer to the edge (i.e.\ the consumer) intuitively seems like
the better approach, an argument can also be made for the inverse
approach. This is because ultimately, the correct decision comes down
to the topology of the network.
\newww{\cref{fig:edge_topology,fig:core_topology}} illustrate two
different network topologies that have very different implications for
the effectiveness of different caching policies.
\cref{fig:edge_topology} shows what we will call an \emph{edge topology}
\ws{where t}here are multiple individual paths from consumers to the
producer that do not intersect at the core of the network, but fray out
at the edge. In such a topology, caching closer to the edge (i.e.\ the
consumers) is beneficial because it will save traversal times from the
edge to the core. Conversely, \cref{fig:core_topology} shows what we
will call a \emph{core topology}\ws{. Here, the} paths between the
consumers and the producer intersect near the producer, with each path
having only one consumer at the edge. In such a topology, it is better
to cache closer to the core (i.e.\ the producer) because that way,
a given cache can serve multiple consumers. \new{Of course, not all
possible topologies will fall neatly in one or the other category; mixed
topologies with a more even distribution of branching paths are also
conceivable.}}

\neww{We can thus see that knowledge of the topology of the given
network can be very helpful when deciding on a caching strategy.
\new{This, in turn, raises the question: What dictates a network's
topology? In \gls*{ICN}, the answer is simple: The topology is directly
informed by the forwarding paths stored in the nodes' \glspl*{FIB}. The
\glspl*{FIB} codify how Interests are forwarded and thus how content is
distributed across the network. Thus, to get a sense of a network's
topology, we need to know how its \glspl*{FIB} are constructed. The
answer to this is not necessarily universal, since \gls*{ICN} enforces
no standards for how \glspl*{FIB} are populated, but in most cases, the
contents of the \glspl*{FIB} will be the direct result of the routing
algorithm used to advertise content. However nodes learn about their
neighbours' contents and in turn inform their neighbours will shape what
their \glspl*{FIB} will look like, which ultimately dictates the entire
network's topology.}}

\neww{The HoPP\new{/Trickle} routing algorithm that we use in this
evaluation makes the individual \glspl*{FIB} organise themselves into
tree topologies based on rank~\cite{gundogan_hopp:_2018}. \neww{This
means} that they \new{resemble} a core topology such as the one shown in
\cref{fig:core_topology}. \neww{Thus, the} paths between consumers and
producers are more likely to cross closer to the producer, making nodes
closer to producers more valuable caching candidates. \new{Our results
in \cref{fig:distance_hops} show that} a strategy like \gls*{LCD}, which
always caches close to the producer, \new{clearly} benefits from this
fact. We can also see that ProbCache-Inv, which has increased caching
probability near the producer, performs \new{a little} better than
default ProbCache with increased probability towards the consumer,
especially at higher distances to source, although \new{this} difference
is not as pronounced.}

\neww{Although Labels does not reach the same hop count reduction as
\gls*{LCD} does, it should be noted that, unlike that strategy, it is
agnostic of topology and thus should be similarly effective in an edge
topology, where we would not expect \gls*{LCD} to reach the same
performance.}

\subsection{Latency and latency reduction}%
\label{sub:distance_latency}

\Cref{fig:distance_latency} shows the average content delivery latency
in relation to the distance \new{to source}. It is immediately obvious
that the latencies for different distances follow the same pattern as
the hop reduction but are slightly more \new{vertically} stretched,
which also follows from the measurements shown in
\cref{fig:hops_latency}: Since the increase in latency with each hop is
linear, the predominant change in latency is solely determined by the
average \new{hops to hit} per distance.

Of course, what we are most interested in is the actual \emph{reduction}
in latency, i.e.\ how much faster we can retrieve content on average
when using a given caching strategy. For this, we compare the average
latency of a given strategy by \new{distance to source} with the average
latency for that number of hops, i.e.\ the expected latency without
caching \new{(cf.~\cref{fig:hops_latency})}. The results are shown in
\cref{fig:latency_reduction_cee}. We can see that \new{the} sweet spot
in terms of latency reduction\ws{,} i.e. the distance at which caching
has the biggest impact, is found at a distance of \rev{4} hops for all
strategies \rev{except \gls*{LCD}, where it is 3}. This was already
\new{implied by} \cref{fig:distance_latency}, where a distance of \rev{4
(3 for \gls*{LCD})} exhibited the first dip in hop count and thus
latency, but when related to \new{the} expected latency it becomes even
more pronounced.

Once again, \gls*{LCD} shows by far the best performance at the peak,
\new{whereas the other strategies are relatively close together. As
might be expected, Labels is once again the strongest contender out of
the remaining strategies, with ProbCache-Inv coming in third.}

\subsection{Overall performance}%
\label{sub:overall_performance}

%\setlength{\tabcolsep}{1em}
%\begin{table}[b]
    %%\centering
    %\footnotesize\sffamily
    %\begin{tabular}{l r r r r}\hline
        %\textbf{Strategy}   & \textbf{Mean}    & \textbf{Mean~~~~~~}      & \textbf{Latency~}   \\
           %& \textbf{hops}    & \textbf{latency~~~~}      & \textbf{reduction}   \\
        %\hline\hline
        %CEE                 & 1.29                  & 22752.08~$\mu$\emph{s}     & 224.75~$\mu$\emph{s}                      \\
        %LCD                 & 1.01                  & 18538.75~$\mu$\emph{s}     & 623.41~$\mu$\emph{s}                      \\
        %Prob(p)             & 1.31                  & 22936.50~$\mu$\emph{s}     & 236.12~$\mu$\emph{s}                      \\
        %ProbCache           & 1.30                  & 23039.65~$\mu$\emph{s}     & 260.96~$\mu$\emph{s}                      \\
        %ProbCache-Inv       & 1.23                  & 22076.50~$\mu$\emph{s}     & 300.20~$\mu$\emph{s}                      \\
        %%Centrality          & 1.13                  & 20122.69~$\mu$\emph{s}     & 572.81~$\mu$\emph{s}                      \\
        %Labels              & 1.18                  & 21394.80~$\mu$\emph{s}     & 336.30~$\mu$\emph{s}                      \\
        %Intervals           & 1.25                  & 22484.31~$\mu$\emph{s}     & 274.96~$\mu$\emph{s}                      \\
                            %&                       &
        
    %\end{tabular}
    %\caption{Mean hops and latencies for the caching strategies~~~}%
    %\label{tab:latencies}
%\end{table}

\begin{figure*}[!t]
\begin{minipage}[b]{0.32\textwidth}
    \centering
    \includegraphics[width=\columnwidth]{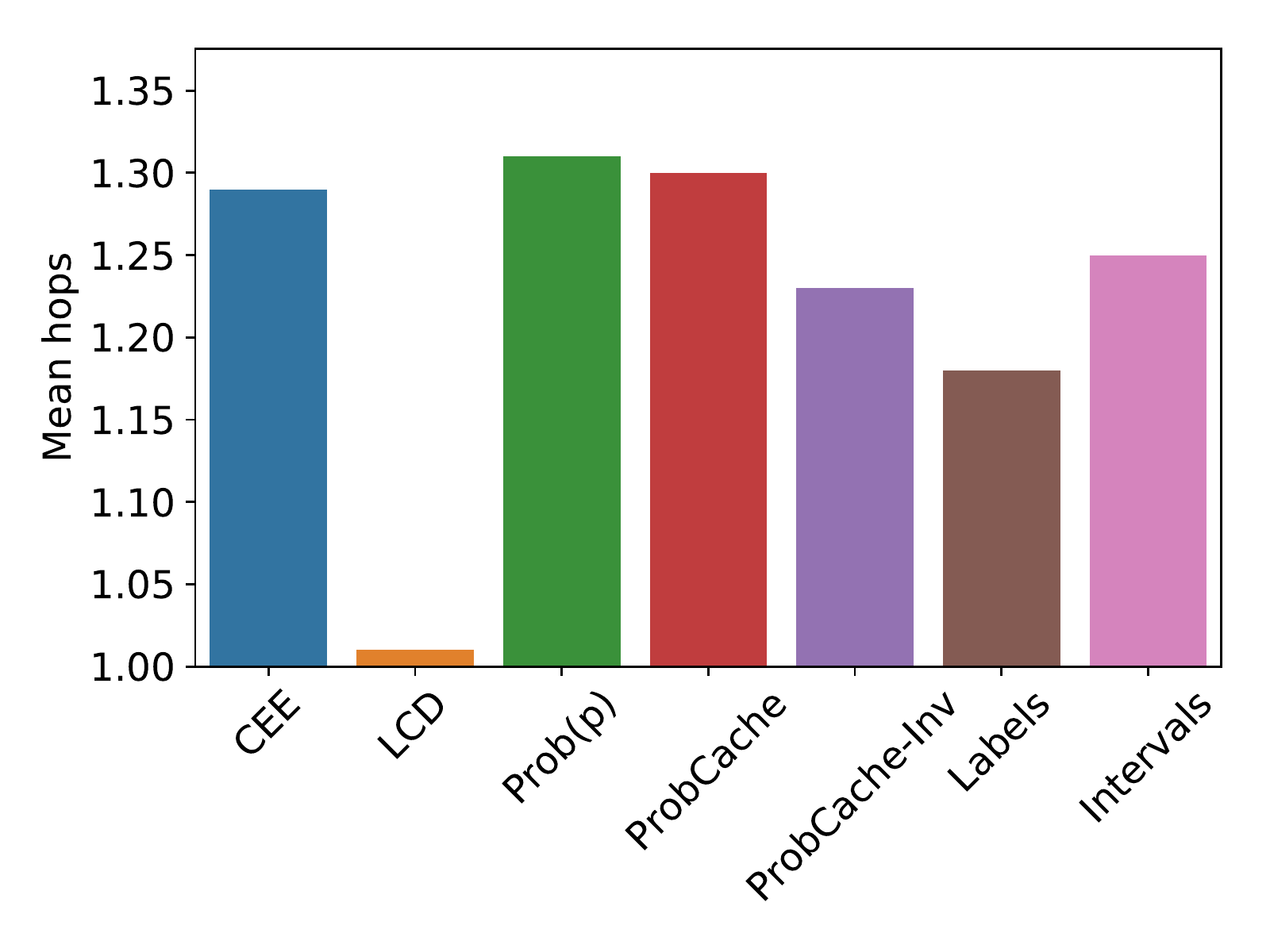}
\caption{Mean hops}%
\label{fig:mean_hops}
\end{minipage}
\hfil
\begin{minipage}[b]{0.32\textwidth}
    \centering
    \includegraphics[width=\columnwidth]{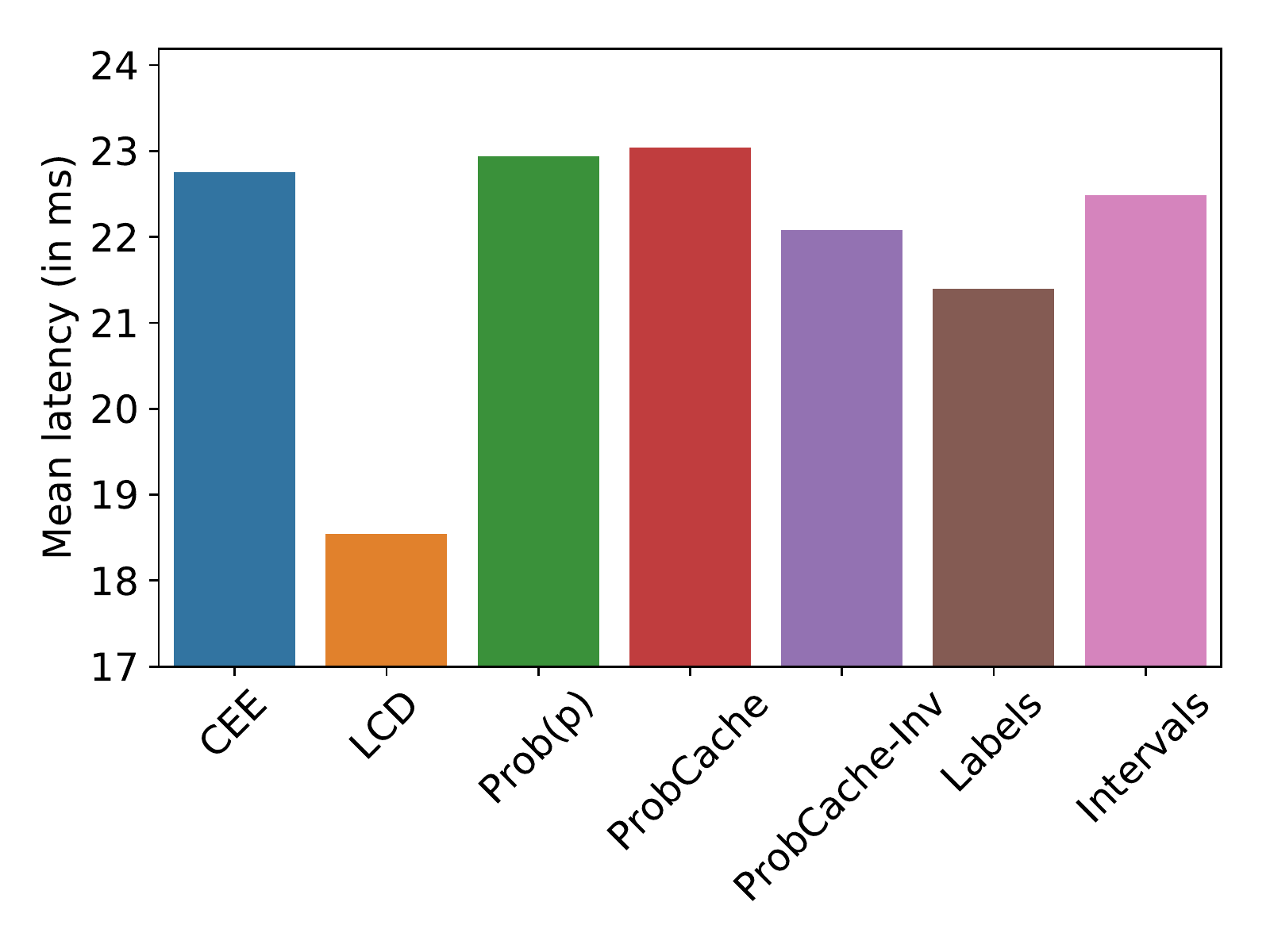}
\caption{Mean latency}%
\label{fig:mean_latency}
\end{minipage}
\hfil
\begin{minipage}[b]{0.32\textwidth}
    \centering
    \includegraphics[width=\columnwidth]{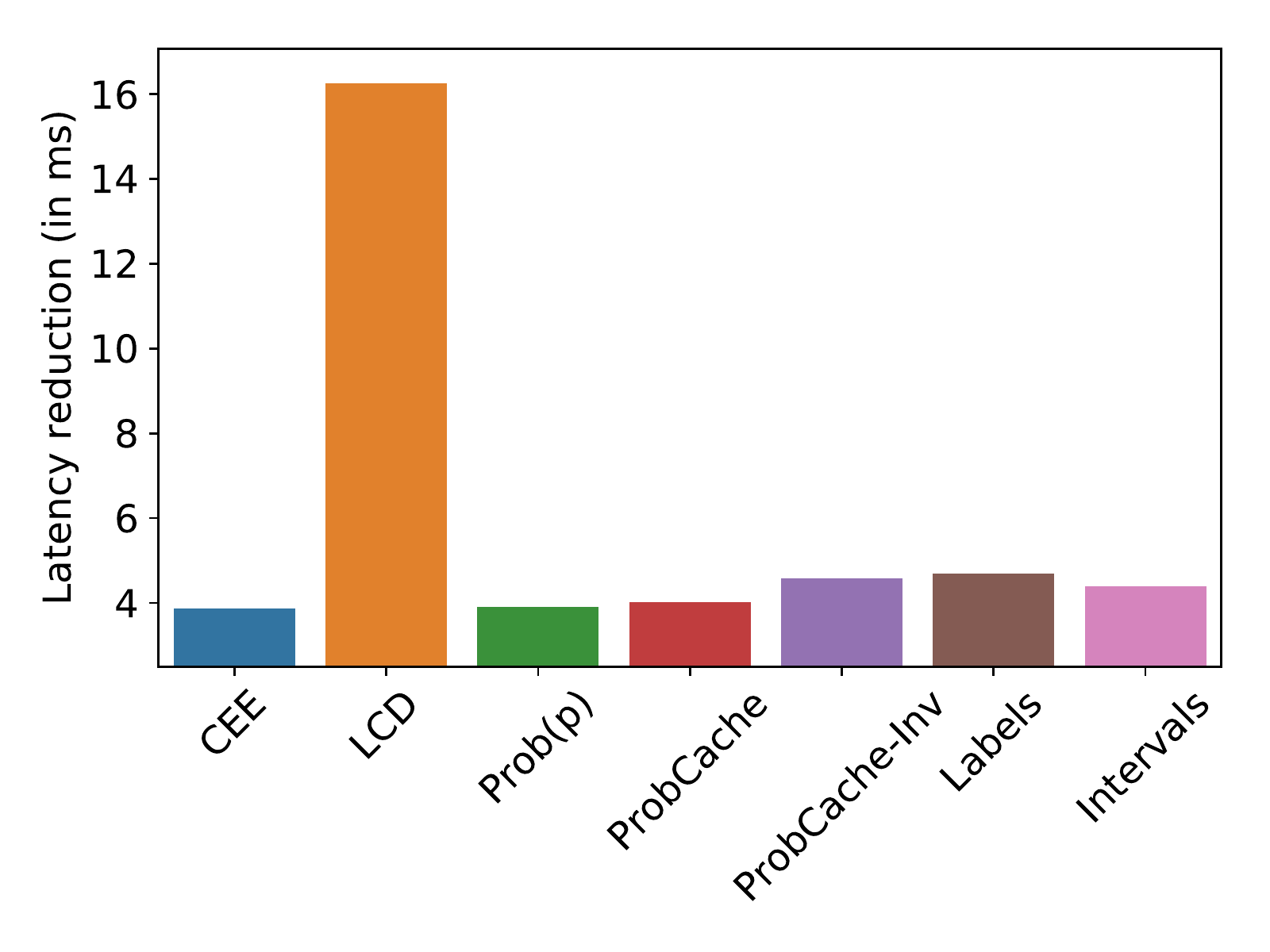}
\caption{Average latency reduction}%
\label{fig:latency_reduction}
\end{minipage}
\end{figure*}

\new{\cref{fig:mean_hops,fig:mean_latency,fig:latency_reduction} show}
an overview of the mean hops, the mean latencies, and the mean reduction
in latency across all path lengths for the different strategies. We can
see that across all strategies, we are able to reduce the mean hops
needed for each Interest to \new{a little over 1}. Although there are no
\new{extreme} outliers, \gls*{CEE}, \new{ProbCache, and Prob(p)} perform
the worst overall, whereas \gls*{LCD} \new{and Labels} perform the best.
Although default ProbCache performs no better than Prob(p), its inverted
variant shows some improvement.

There are several conclusions we can draw from this. The first is that
as described above, due to the fact that \wsq{our chosen routing
algorithm results in a core topology,} 
%\marginpar{\scriptsize\hspace*{-4ex}\ask{\parbox[]{13ex}{\raggedright The fact that the routing algorithm results in a core topology wasn't obvious. Since it is mentioned now, one would then question the need to also study the edge topology scenario.}}}
caching near the producer is the more effective policy, making
\gls*{LCD} a good choice. From the results, we can also see that
strategies that do not take elements of the topology into account (i.e.\
\gls*{CEE} and \new{Prob(p)}) generally do not perform as well as those
that do\new{, although Labels goes against that trend}. The middle
ground is covered by ProbCache-Inv, which does consider topology but
still introduces an element of chance. A tentative conclusion from this
is that deterministic, topology-based approaches provide better content
placement than probabilistic ones.

%!TEX root = paper.tex
%==============================================================================

\section{Related Work}%
\label{sec:related_work}

In this section, we will give an overview of existing comparative
\gls*{ICN} caching studies \ws{with a focus on} \gls*{IoT}.
\new{Although extensive research has been conducted into caching in
\gls*{ICN}, there have been comparatively few studies on caching as it
relates to information-centric \gls*{IoT}.}

\new{There have been multiple surveys~\cite{zhang_caching_2013,
fang_survey_2014, zhang_survey_2015} on caching strategies for
traditional \gls*{ICN}, as well as a number of comparative
studies~\cite{tarnoi_performance_2014, zhang_survey_2015}.
Carofiglio~\emph{et~al.} have produced a body of work~\cite{LCN2015:LAC,
carofiglio_focal:_2015, carofiglio_analysis_2016, carofiglio_joint_2016}
focusing on latency effects in traditional \gls*{ICN} caching, but their
solutions do not address the idiosyncrasies of \gls*{IoT} environments.
Caching research focused specifically on information-centric \gls*{IoT}
has not had much exposure at the time of writing.}

\new{The most comprehensive overview of existing ICN-IoT caching schemes
was performed by
Arshad~\emph{et~al.}~\cite{arshad_information-centric_2017,
arshad_recent_2018}. However, this is a purely qualitative survey, and
no experimental evaluation of the strategies was performed. The first
comparative studies on \gls*{ICN} caching strategies specifically in the
\gls*{IoT} were presented by
Hail~\emph{et~al.}~\cite{hail_performance_2015} and
Meddeb~\emph{et~al.}~\cite{meddeb_how_2017}. They used simulated
environments to compare several cache decision and cache replacement
policies. For our own previous work~\cite{pfender_performance_2018}, we
took some inspiration from these studies while showcasing a greater
number of caching decision and cache replacement strategies and using
more performance metrics for our evaluation; we also performed all of
our experiments on physical \gls*{IoT} hardware operating in realistic
conditions. An important conclusion from our previous study was that
cache replacement policies have little to no impact on the performance
of ICN-IoT in-network caching, which is why in this study, we focus
exclusively on caching decision policies.}

\new{To the best of our knowledge, no previous study has explicitly
taken the effects of the routing algorithm and the network topology into
account when evaluating in-network caching strategies for
information-centric \gls*{IoT}, and no other comparative study apart
from our own previous research has been performed using physical
\gls*{IoT} hardware.}

%!TEX root = paper.tex
%==============================================================================

\section{Conclusions and Future Work}%
\label{sec:conclusion}

We have presented a comparison and evaluation of several different
caching strategies for information-centric \gls*{IoT}, focusing on their
effects on content delivery latency. Our results \newww{indicate} that
the topology of the network has \newww{an} impact on which
cache distribution is optimal, and thus \newww{it may be fruitful to use
caching algorithms that are optimised for the given topology}. However,
caching strategies that ignore topology in favour of other approaches,
such as stratifying the data evenly across the network, are also
promising and may be preferable if the topology is unknown, mutable, or
a hybrid between different types. \newww{The} network topology is
influenced by the choice of routing algorithm, \new{thereby} creating
a direct link between routing algorithm and caching strategy. In other
words, a holistic caching solution for information-centric \gls*{IoT}
should take \ws{all of these} aspects into account.

\newww{Of course, determining} the \newww{optimal} caching strategy for
a \newww{given} \gls*{IoT} application \newww{also} depends on the
requirements of that particular application as well as the
\newww{constraints imposed by the hardware}. More computationally
intensive caching strategies may yield better results, but may take too
\ws{much} resources away from the actual application. Overall, however,
\newww{our results indicate} that even simple caching strategies
\new{such as \gls*{LCD} or Labels} can improve performance compared to
\new{indiscriminate caching}, so they will almost always be worth
considering.

The direct link between the choice of routing algorithm and the
effectiveness of the caching strategy is a phenomenon that, to the best
of our knowledge, has not been explicitly investigated in \gls*{ICN}
caching research to date. Most research that compares caching strategies
does not take the routing algorithm into account at all; often,
\glspl*{FIB} are just assumed to be populated \emph{a priori}. We
believe that an in-depth comparison of caching strategy performance on
topologies created by different routing algorithms would yield valuable
insights and pave the way towards a holistic solution. Similarly, future
research should focus more strongly on different topologies likely to be
encountered in \gls*{IoT} deployments and how they affect the choice of
caching strategy. \neww{Therefore, our future work will have the aim of
comparing the effectiveness of different caching strategies in core
topologies as well as edge topologies, and work towards caching
solutions that are effective regardless of topology.}

\ifCLASSOPTIONcaptionsoff
  \newpage
\fi



\bibliographystyle{IEEEtran}
\bibliography{NDN}
%\begin{thebibliography}{1}

%H.~Kopka and P.~W. Daly, \emph{A Guide to {\LaTeX}}, 3rd~ed.\hskip 1em plus
  %0.5em minus 0.4em\relax Harlow, England: Addison-Wesley, 1999.

%\end{thebibliography}


\begin{IEEEbiography}[{\includegraphics[width=1in,height=1.25in,clip,keepaspectratio]{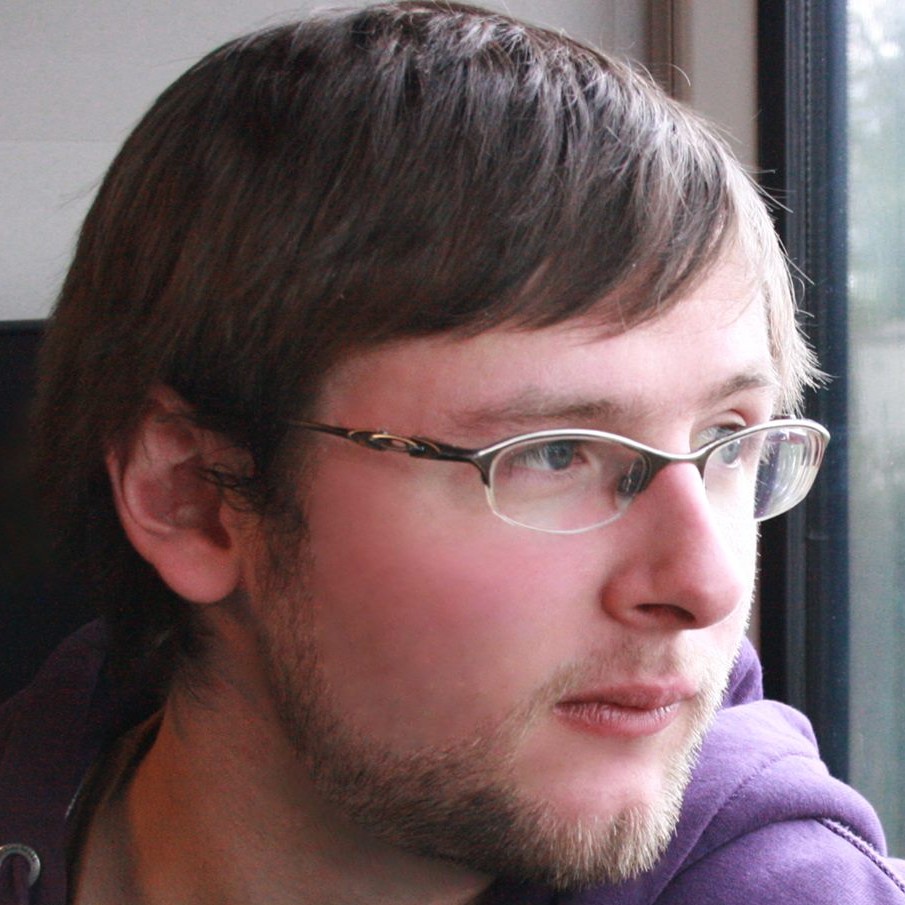}}]{Jakob Pfender}
received his B.Sc.\ and M.Sc.\ degrees in computer science from Freie
Universität Berlin, Germany, in 2013 and 2016. He is currently pursuing
a PhD degree in network engineering at Victoria University of
Wellington, New Zealand. His research interests include new protocols
for the Internet of Things, resource-constrained Information-Centric
Networking, indoor localisation, and distributed event detection.
\end{IEEEbiography}

\begin{IEEEbiography}[{\includegraphics[width=1in,height=1.25in,clip,keepaspectratio]{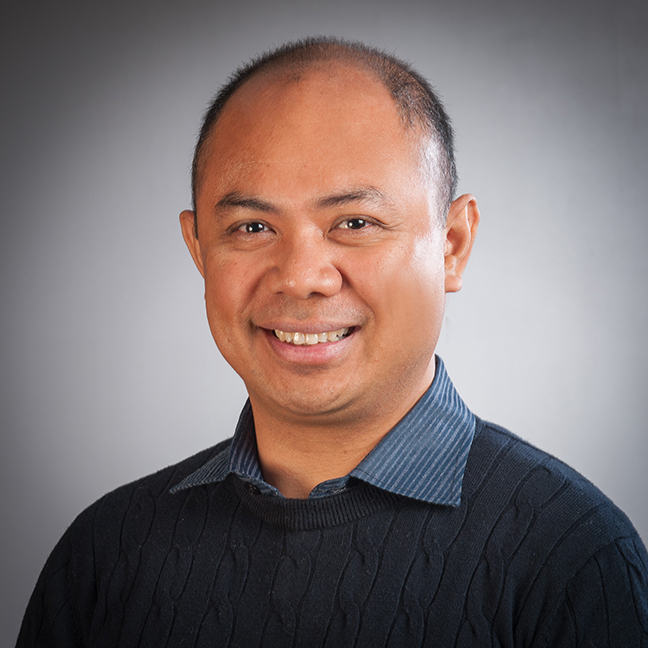}}]{Alvin C. Valera}
is currently a Senior Lecturer at the School of Engineering and Computer
Science, Victoria University of Wellington. He obtained the Bachelor of
Science degree (computer engineering) from the University of the
Philippines on 1998, Master of Science (computer science) and Ph.D.
(electrical and computer engineering) from the National University of
Singapore in 2003 and 2015, respectively. His current research interests
interests are in the development of protocols and architectures for the
Internet of Things (IoT) and resource-constrained information-centric
networks.
\end{IEEEbiography}

\begin{IEEEbiography}[{\includegraphics[width=1in,height=1.25in,clip,keepaspectratio]{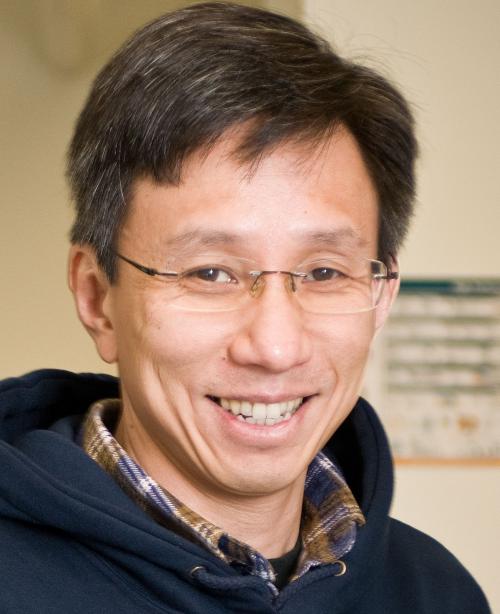}}]{Winston K.G. Seah}
received the Dr.Eng.\ degree from Kyoto University, Kyoto, Japan, in
1997. He is currently a Professor of network engineering with the School
of Engineering and Computer Science, Victoria University of Wellington,
New Zealand. Prior to this, he worked for over 16 years
in mission-oriented industrial research, taking ideas from theory to
prototypes, most recently as a Senior Scientist with the Institute for
Infocomm Research, Singapore. His current research interests include
Internet of Things, wireless sensor networks powered by ambient energy
harvesting, wireless multihop networks, software-defined networking, and
5G access protocols for machine-type communications.
\end{IEEEbiography}


%\newpage



%\vfill


\end{document}